\documentstyle[12pt]{article}

\begin{document}

\begin{flushright}
IMSc/2006/10/22
\end{flushright} 

\vspace{2mm}

\vspace{2ex}

\begin{center}
{\large \bf 

A Principle to Determine the Number 

\vspace{2ex}

($3 + 1$) of Large Spacetime Dimensions 

} 

\vspace{8ex}

{\large  S. Kalyana Rama}

\vspace{3ex}

Institute of Mathematical Sciences, C. I. T. Campus, 

Tharamani, CHENNAI 600 113, India. 

\vspace{1ex}

email: krama@imsc.res.in \\ 

\end{center}

\vspace{6ex}

\centerline{ABSTRACT}
\begin{quote} 

We assume that our universe originated from highly excited and
interacting strings with coupling constant $g_s = {\cal O} (1)$.
Fluctuations of spacetime geometry are large in such strings and
the physics dictating the emergence of a final spacetime
configuration is not known. We propose that, nevertheless, it is
determined by an entropic principle that the final spacetime
configuration must have maximum entropy for a given amount of
energy. This principle implies, under some assumptions, that the
spacetime configuration that emerges finally is a $(3 + 1)$ --
dimensional FRW universe filled with $w = 1$ perfect fluid and
with $6$ -- dimensional compact space of size $l_s$; in
particular, the number of large spacetime dimensions is $d = 3 +
1$. Such an universe may evolve subsequently into our universe,
perhaps as in Banks -- Fischler scenario.

\end{quote}

\vspace{2ex}


\newpage

{\bf 1.}  
In superstring theory the number of critical dimensions of the
spacetime is $9 + 1$ whereas the spacetime in the observed
universe is $3 + 1$ dimensional. It is expected that $6$ spatial
dimensions will compactify to string size by some mechanism,
resulting in a $3 + 1$ -- dimensional universe. One proposal for
such a mechanism \cite{bv} involves winding modes of the strings
and other recent ones \cite{durrer} involve various $D$ --
branes. \footnote{ In the context of $11$ -- dimensional
supergravity with Freund -- Rubin type compactifications, it is
shown using quantum cosmology techniques that observed
spacetime must be $4$ -- dimensional \cite{zwu}. }

Consider an FRW universe. In string theory, it is described
using low energy effective action for zero modes. In the past,
as time decreases, the temperature of the universe eventually
reaches the string scale at which the universe is to be
described by stringy variables. In this context a stringy
correspondence principle, analogous to that of Horowitz and
Polchinski \cite{hp, s}, is formulated for the evolution of the
state of the universe \cite{k}. Briefly, according to this
principle, at temperatures lower than string scale the universe
state evolves as in FRW cosmology whereas at higher temperatures
it evolves as highly excited strings. At the transition, the
entropies and energies in these two descriptions differ by
numerical factors of ${\cal O}(1)$ as shown in \cite{k} in the
weak coupling limit where the string coupling constant $g_s \ll
1$.

This transition can be thought of as the universe turning into
highly excited strings or, in reverse which is not well
understood, as highly excited strings turning into an expanding
FRW universe. Here, we assume that a similar picture holds for
our universe also where the string coupling constant $g_s =
{\cal O} (1)$. Accordingly then, as one follows the evolution in
the past, our universe turns into highly excited and, since $g_s
= {\cal O} (1)$, highly interacting strings; conversely, our
universe originated from such highly excited and interacting
strings.

However, in a system composed of highly excited strings with a
large number of interacting degrees of freedom, fluctuations of
spacetime geometry are large and associated spacetime concepts
are not well defined. These fluctuations may be pictured as
spacetime configurations emerging and dissolving back into
strings from time to time. Finally a spacetime configuration,
stable against dissolving back, must emerge leading to our
universe. This is clearly a necessary requirement for the
assumption here that our universe, where spacetime is well
defined, originated from highly excited and interacting strings.

The physics dictating the course of the emergence of such a
final configuration is beyond our grasp. In this paper we
propose that, nevertheless, it is determined by an entropic
principle that the final spacetime configuration must have
maximum entropy for a given amount of energy. This is a standard
principle generically applicable for systems containing a large
number of interacting degrees of freedom.

We then study a consequence of the entropic principle and show
that it implies, under a few further assumptions, that the
spacetime configuration that emerges finally is a $(3 + 1)$ --
dimensional FRW universe filled with $w = 1$ perfect fluid
\footnote{ which can also be thought of as black hole fluid
\cite{bf} } and with $6$ -- dimensional compact space of size
$l_s$; in particular, the number of large spacetime dimensions
is $d = 3 + 1$. Such an universe may be taken to evolve
subsequently into our universe, perhaps as in Banks -- Fischler
scenario \cite{bf}.

This paper is organised as follows. In section {\bf 2} we
present the relevant expressions for an FRW universe. In section
{\bf 3} we present our proposal and study a consequence of the
entropic principle. In section {\bf 4} we conclude with a few
remarks.

\vspace{4ex}

{\bf 2.}  
Consider the evolution of a $d$ -- dimensional spatially flat
FRW universe containing a perfect fluid with density $\rho$ and
pressure $p = w \rho$. We assume that such an universe
originated from a ten dimensional superstring theory
compactified on a $p = (10 - d)$ -- dimensional compact space.
With $\hbar = c = 1$, the $d$ -- dimensional Planck length
$l_{pl}$ is then given by
\begin{equation}\label{pltos}
l_{pl}^{d - 2} \simeq \frac{g_s^2 l_s^8}{V_p}
\end{equation}
where $g_s$ is the string coupling constant and $ = {\cal O}(1)$
for our universe, $l_s$ is the string length, $V_p$ is the
volume of the compact space and, here and in the following,
$\simeq$ denotes that numerical factors of ${\cal O}(1)$ are
omitted. If the compact space is {\em e.g.}  toroidal with sizes
$L_1, \cdots, L_p$ then $V_p \simeq \prod_1^p L_i$.

The parameters of this FRW universe are $w$, $V_p$, and $d$ and
their ranges are restricted. The standard energy conditions
imply that $- 1 \le w \le 1$. For toroidal compactification,
T-duality symmetry of the string theory implies that $V_p
\stackrel{>} {_\sim} l^p_s$. We assume this to be the case in
general also. In superstring theory $d \le 10$. Also, gravity
plays an important role in what follows and, hence, we further
assume that $d \ge 4$ since gravity is not a propagating degree
of freedom in lower dimensions.

The relevant line element $d s$ is given, in the standard
notation, by
\begin{equation}\label{dsd}
d s^2 = - d t^2 + a^2 \left( d r^2 + r^2 d \Omega_{d - 2}^2
\right) \; . 
\end{equation}
Solving the equations of motion one gets, with $\alpha =
\frac{2}{(d - 1) (1 + w)}$,
\begin{equation}\label{soln}
\frac{a(t)}{a_{pl}} = \left( \frac{t}{l_{pl}} \right)^\alpha
\; \; , \; \; \; 
\rho(t) \simeq \frac{1}{l_{pl}^d} \;
\left( \frac{a_{pl}}{a} \right)^{(d - 1) (1 + w)} \; .
\end{equation} 
The constant $w \le 1$. For example, $w = \frac{1} {d - 1}$ for
radiation field whereas $w = 1$ for a massless scalar field.
Such fields are present in string theory. When the universe
contains many perfect fluids with different $w's$ then that with
the highest value of $w$ dominates in the past when the
temperature $T$ is high.
  
It is natural to take the size of the universe to be given by
the size of its horizon which encompasses the maximum region
within which causal contact is possible. We do so in the
following. The entropy $S$ and the energy $E$ of the universe
can then be defined to be those contained within its horizon and
are given by
\begin{equation}\label{ser}
S = \sigma \; V_{d - 1} r_H^{d - 1} 
\; \; , \; \; \; 
E = \rho \; V_{d - 1} L_H^{d - 1} 
\end{equation}
where $\sigma = \frac{(\rho + p)}{T} \; a^{d - 1}$ is the
constant comoving entropy density, $V_n$ is the volume of an
unit $n$ -- dimensional ball, and 
\[
r_H = \int_0^t \frac{d t}{a} = \frac{l_{pl}}{(1 - \alpha) a_{pl}} \; 
\left( \frac{t}{l_{pl}} \right)^{1 - \alpha} 
\; \; , \; \; \; 
L_H = r_H a = \frac{t}{1 - \alpha} 
\]
are the comoving coordinate and the physical size of the horizon
respectively. Written in terms of $t$, the expressions for $S$
and $E$ become
\begin{equation} 
S =  C_S \left( \frac{t}{l_{pl}} 
\right)^{(d - 1) (1 - \alpha)} 
\; \; , \; \; \; \; 
E =  \frac{C_E}{l_{pl}} \left( \frac{t}{l_{pl}} 
\right)^{d - 3}
\end{equation}
where $C_S$ and $C_E = {\cal O}(1)$ are numerical coefficients.
Holographic principle \cite{fs} implies that $C_S = {\cal O}(1)$
also. See \cite{k} more detailed expressions. The entropy $S$ as
a function of energy $E$ is then given, with $b = \frac{d - 3 +
(d - 1) w} {(d - 3) (1 + w)}$, by
\begin{equation}\label{se} 
S(E) \simeq \left( l_{pl} E \right)^b
\simeq \left( \frac{g_s^2 l_s^8}{V_p} 
\right)^{\frac{b}{d - 2}} \; E^b \; \; .
\end{equation}

\vspace{4ex}

{\bf 3.}  
The FRW description of the universe given above is obtained
using low energy effective action for string zero modes. In the
past, as time decreases, the temperature of the universe
increases and reaches the string scale $\simeq \frac{1}{l_s}$.
At such a scale a large number of higher modes of strings are
excited and their effects must be included. The FRW description
of the universe given above is then to be replaced by a stringy
description.

In this context, a stringy correspondence principle is
formulated for the evolution of the state of the universe
\cite{k}, in analogy with that of Horowitz and Polchinski for
black hole states \cite{hp, s}. According to this principle,
there is a correspondence between a FRW universe state and a
highly excited string state. When the temperature is lower than
string scale the universe state evolves as in FRW cosmology and
is described by FRW variables. When the temperature is higher
than string scale the universe state evolves as highly excited
strings and is to be described by stringy variables. At the
transition, the entropies and energies in these two descriptions
differ by numerical factors of ${\cal O}(1)$.

This is shown for the transition from FRW description to the
stringy one in the weak coupling limit where $g_s \ll 1$. The
transition can be thought of as the universe turning into highly
excited strings as one follows its evolution in the past. The
reverse transition, namely highly excited strings turning into
an expanding FRW universe, is not well understood. See \cite{k}
for a few remarks on this transition and \cite{hp2} for a
detailed study of the transition from strings to black hole.

Here, we assume that a similar picture holds for our universe
also where the string coupling constant $g_s = {\cal O}(1)$.
Accordingly then, as one follows the evolution in the past, our
universe turns into highly excited and, since $g_s = {\cal O}
(1)$, highly interacting strings; conversely, our universe
originated from such a highly excited and highly interacting
strings.

The dynamics of such highly excited and interacting strings are
difficult to study at present. Some of the difficulties are that
all excited modes and their non trivial interactions must be
included; the standard low energy effective actions are not
applicable; fluctuations of spacetime geometry, described
essentially by string zero modes, are large and associated
spacetime concepts are not well defined.

These spacetime fluctuations may be pictured as some spacetime
configuration emerging and after some time dissolving back into
strings, then some other configuration emerging and dissolving
back, and so on. A spacetime configuration here is to be
parametrised by {\em e.g.} the number and sizes of spacetime
dimensions, constituent fields in the spacetime, etcetera. These
parameters are different for different configurations. Finally a
spacetime configuration emerges from strings which is stable
against dissolving back, and whose subsequent evolution,
described by an effective action, will lead to our universe.
\footnote{ This picture is loosely analogous to a solid
structure emerging from a liquid.}  That such a final
configuration, with well defined spacetime concepts, must emerge
from strings is clearly a necessary requirement for the
assumption here that our universe, where spacetime is well
defined, originated from highly excited and interacting strings.

Although the physics dictating the course of the emergence of
such a final configuration is beyond our grasp, it may be
possible to determine the final configuration itself. We propose
that it is determined by the following entropic principle:
\begin{quote}
{\em The spacetime configuration that emerges finally from
highly excited and interacting strings is the one which has
maximum entropy for a given amount of energy.}
\end{quote}
Here and in the following it is assumed that energy $E \gg
\frac{1}{l_s}$.

This is the standard entropic principle \footnote{ In
statistical mechanics where entropic principle is commonly
applied, if microcanonical and canonical ensembles are
equivalent then maximising entropy translates into minimising
free energy; otherwise microcanonical ensemble, and therefore
maximising entropy, is more fundamental.} and is applicable if
the system is sufficiently ergodic and mixing so that various
configurations can be sampled and the maximum entropic one
picked out. This is generically the case if the system contains
a large number of interacting degrees of freedom. The present
case of highly excited and interacting strings with coupling
constant $g_s = {\cal O}(1)$ is indeed a system with a large
number of interacting degrees of freedom. Hence, it is likely to
be sufficiently ergodic and mixing so that entropic principle
can be applied to it. Assuming this to be the case, we now apply
the entropic principle and study a consequence.

Assume that a $d$ -- dimensional FRW universe described earlier
is the spacetime configuration that emerges finally from the
strings whose coupling constant $g_s = {\cal O}(1)$ and string
length is $l_s$. For a given energy $E$, the entropy $S(E)$ of
the universe is given in equation (\ref{se}). The parameters of
this spacetime are $w$, $V_p$, and $d$ and their ranges are
restricted. The standard energy conditions imply that $w \le 1$.
As per our assumptions, the T-duality symmetry of the string
theory implies $V_p \stackrel{>} {_\sim} l_s^p$; and $d \ge 4$
since gravity is not a propagating degree of freedom in lower
dimensions. According to the entropic principle, the values of
these parameters will be such as to maximise $S(E)$ for a given
$E$. It can be easily seen that the values which maximise $S(E)$
are
\[
w = 1 \; \; , \; \; \; 
V_p \simeq l_s^p \; \; , \; \; \; 
d = 4 \; 
\]
and that the maximum entropy $S_{max}(E)$ is 
\begin{equation}\label{smax}
S_{max}(E) \; \simeq \; l^2_{pl} E^2 \; 
\simeq \; \; g^2_s l^2_s E^2 \; .
\end{equation}
For open or closed FRW universe, or for an universe containing
other perfect fluids also such as radiation ($w = \frac{1}{d -
1}$) or a cosmological constant ($w = - 1$), the $w = 1$ perfect
fluid dominates the universe in the past. Therefore, to the
leading order, its entropy $S(E)$ is still given by that for $w
= 1$ fluid, namely by equation (\ref{se}) with $w = 1$. Hence,
the entropic principle can not determine whether the universe
that emerges finally is flat, open, or closed or the details of
its subleading matter contents.

In reference \cite{bf}, Banks and Fischler present a detailed
scenario where an universe filled with $w = 1$ perfect fluid,
also thought of as black hole fluid, \footnote{ Entropic
principle applied to $d$ -- dimensional black holes also leads
to $V_p \simeq l_s^p$ and $d = 4$.} evolves into radiation
dominated FRW universe whose subsequent evolution proceeds as in
the standard cosmology. Perhaps then the above $(3 + 1)$ --
dimensional FRW universe with $w = 1$ perfect fluid, which is
similar to the one studied in \cite{bf} but is obtained here as
a consequence of entropic principle, may also be taken to evolve
as in Banks -- Fischler scenario into radiation domianted
universe and then subsequently into our universe.

Note that gravity plays an important role. In the presence of
gravity, a space filled with {\em e.g.} radiation fluid ($w =
\frac{1}{d - 1}$) evolves as an FRW universe, with the entropy
$S$ and the energy $E$ contained within horizon related as in
equation (\ref{se}). In the absence of gravity, the entropy $S$
and the energy $E$ of radiation fluid are related as
\[
S(E) \simeq \left( L E \right)^{\frac{d - 1}{d}}
\]
where $L$ is the spatial size. Entropic principle, {\em i.e.}
maximising $S(E)$ with respect to the parameters $d \le 10$ and
$L$, would then give $d = 10$ and $L \to \infty$. This may
perhaps be the case for free, or weakly interacting, strings
where the coupling constant $g_s = 0$, or $g_s \ll 1$, but
applying entropic principle in this context is questionable
because the interactions are absent, or arbitrarily weak.

We are not aware of any other spacetime configuration \footnote{
The obvious case of black hole fluid can be thought of as
equivalent to $w = 1$ fluid \cite{bf}.} where gravity is present
and whose entropy $S(E)$ is greater than $S_{max}(E)$ given in
equation (\ref{smax}). If we assume that no such configuration
exists then it follows that the entropic principle implies that
the spacetime configuration that emerges finally from a highly
excited and interacting strings is a $(3 + 1)$ -- dimensional
FRW universe filled with $w = 1$ perfect fluid and with $6$ --
dimensional compact space of size $l_s$. Such a spacetime would
then evolve into our universe, perhaps as in Banks -- Fischler
scenario. In particular, the entropic principle would thus have
determined the number, $d = 3 + 1$, of large spacetime
dimensions.

\vspace{4ex}

{\bf 4.} 
We have proposed here that the spacetime configuration that
emerges finally from highly excited and interacting strings is
determined by the entropic principle. This is a standard
principle and is generically applicable for systems containing a
large number of interacting degrees of freedom. 

Highly excited and interacting strings is a system with a large
number of interacting degrees of freedom justifying thereby the
application of entropic principle. The interaction effects are
strong since $g_s = {\cal O}(1)$ and one may therefore expect
that various spacetime configurations are sampled efficiently;
that lower entropic configurations are short lived; and that the
spacetime configuration which emerges finally is one of maximum
entropy. The entropic principle implies, under the assumptions
mentioned earlier, that the spacetime configuration that emerges
finally is a $(3 + 1)$ -- dimensional FRW universe filled with
$w = 1$ perfect fluid and with $6$ -- dimensional compact space
of size $l_s$; in particular, the number of large spacetime
dimensions is $d = 3 + 1$.

However, the entropic principle does not give the details of the
dynamical features involved such as how the spacetime
configurations emerge from and dissolve back into strings; how
various configurations are sampled; the `life time' of lower
entropic configurations; or, the time scale over which the
maximum entropic configuration emerges finally.

One may try to understand such details in the weak coupling
limit where $g_s \ll 1$ and where perturbative techniques may be
applied. But, such an understanding may not be possible because
one usually assumes a background spacetime in a perturbative
formulation and also because the interactions are arbitrarily
weak when $g_s \ll 1$ and applying entropic principle may then
be questionable. One will then need techniques applicable when
$g_s = {\cal O} (1)$.




\end{document}